\documentclass[pra,twocolumn,superscriptaddress]{revtex4-1}

\usepackage[utf8]{inputenc}
\usepackage[english]{babel}
\usepackage[T1]{fontenc}
\usepackage{lmodern}

\usepackage{graphicx}
\usepackage{subfigure}

\usepackage{amsmath,amsthm,amsfonts,amssymb}
\usepackage{dsfont}

\usepackage{microtype}

\usepackage[breaklinks=true,colorlinks]{hyperref}

\newcommand{\coloneq}{\mathrel{\mathop:}=}
\newcommand{\eqcolon}{=\mathrel{\mathop:}}
\newcommand{\Tr}{\operatorname{Tr}}
\newcommand{\ket}[1]{\left|{#1}\right\rangle}

\newcommand{\ketbra}[2]{\left|{#1}\middle\rangle\middle\langle{#2}\right|}
\newcommand{\proj}[1]{\ketbra{#1}{#1}}

\newcommand{\kB}{k_\mathrm{B}}

\usepackage{array}
\newcolumntype{P}[1]{>{\centering\arraybackslash}p{#1}}

\begin{document}

\title{Collectively enhanced thermalization via multiqubit collisions}

\author{Angsar Manatuly}
\affiliation{Department of Physics, Ko\c{c} University, 34450 Sar{\i}yer, \.{I}stanbul, Turkey}

\author{Wolfgang Niedenzu}
\affiliation{Institut f\"ur Theoretische Physik, Universit\"at Innsbruck, Technikerstra{\ss}e~21a, A-6020~Innsbruck, Austria}

\author{Ricardo Rom\'an-Ancheyta}
\affiliation{Department of Physics, Ko\c{c} University, 34450 Sar{\i}yer, \.{I}stanbul, Turkey}

\author{Bar{\i}\c{s} \c{C}akmak} 
\affiliation{Department of Physics, Ko\c{c} University, 34450 Sar{\i}yer, \.{I}stanbul, Turkey}
\affiliation{College of Engineering and Natural Sciences, Bah\c{c}e\c{s}ehir University, Be\c{s}ikta\c{s}, Istanbul 34353, Turkey}

\author{\"{O}zg\"{u}r E. M\"{u}stecapl{\i}o\u{g}lu}
\email{omustecap@ku.edu.tr}
\affiliation{Department of Physics, Ko\c{c} University, 34450 Sar{\i}yer, \.{I}stanbul, Turkey}
  
\author{Gershon Kurizki} 
\affiliation{Department of Chemical Physics, Weizmann Institute of Science, Rehovot 7610001, Israel}

\date{May 2, 2019}

\begin{abstract}
  We investigate the evolution of a target qubit caused by its multiple random collisions with $N$-qubit clusters. Depending on the cluster state, the evolution of the target qubit may correspond to its effective interaction with a thermal bath, a coherent (laser) drive, or a squeezed bath. In cases where the target qubit relaxes to a thermal state its dynamics can exhibit a quantum advantage, whereby the target-qubit temperature can be scaled up proportionally to $N^2$ and the thermalization time can be shortened by a similar factor, provided the appropriate coherence in the cluster is prepared by non-thermal means. We dub these effects quantum super-thermalization due to its analogies to super-radiance. Experimental realizations of these effects are suggested.
\end{abstract}

\maketitle

\section{Introduction}

Heat flow or exchange between systems is an abundant resource, yet its control is subtle, particularly in the quantum domain~\cite{kosloff2014quantum,gelbwaser2015thermodynamics,goold2016role,vinjanampathy2016quantum}. A major motivation for the study of such control is the quest for quantum advantage in the exploitation of this resource for useful purposes, e.g., in heat engines~\cite{alicki1979quantum,kosloff1984quantum,scully2003extracting,quan2007quantum,abah2012single,kosloff2014quantum,gelbwaser2015thermodynamics,uzdin2015equivalence,goold2016role,rossnagel2016single,vinjanampathy2016quantum,klaers2017squeezed,klatzow2019experimental,guff2018power} or heat diodes~\cite{ordonez2017quantum,kargi2019quantum,ronzani2018tunable}. A prerequisite for such studies is our ability to control thermalization. According to the open-system approach, thermalization occurs whenever one can decompose the Hilbert space into a small subspace (``system'') and a much larger subspace (``bath''), assuming their interaction is weak enough to be treated perturbatively~\cite{breuerbook}. It is much less clear what are the criteria for thermalization in system-bath complexes subject to nonclassical state preparation, driving or measurements, even if they are describable by a master equation~\cite{breuerbook,scullybook}. Such is the case of the micromaser setup~\cite{filipowicz1986theory} wherein the cavity mode (the ``system'') sequentially and randomly interacts with a ``bath'' composed of quantum objects (atoms~\cite{filipowicz1986theory} or atomic clusters~\cite{dag2016multiatom,dag2019temperature}) that are traced out after each interaction~\cite{scullybook}.

\par

Here we examine the issue of thermalization and heat exchange for a composite system consisting of a target qubit (the system) that randomly, repeatedly, interacts with a cluster of $N$ \emph{identical} spin-$1/2$ (qubit) particles (the bath). In such a composite system asymptotic thermalization or its absence turns out to be strongly dependent on the initial state of the $N$-qubit cluster: We allow for \emph{multi-qubit coherences} in the cluster bath and show that depending on the type of coherences the bath may drive the target qubit into either a thermal or a coherently-displaced state. Even in cases where the target qubit thermalizes, it can exhibit a \emph{quantum advantage}, whereby the target-qubit temperature can be scaled up with $N^2$, and the thermalization time can be shortened by a similar factor, provided the appropriate coherence in the cluster is prepared by non-thermal means. We dub these effects quantum super-thermalization because of their similarity (but also differences) with superradiance. Experimental realizations of these effects are suggested.

\par

In addition to the insights provided by this study into thermalization and non-thermalization conditions for composite quantum systems, the controllability of their thermalization rates by subspace-state preparation has potential technological significance that has only sporadically been explored so far: heat engines~\cite{alicki1979quantum,kosloff1984quantum,scully2003extracting,quan2007quantum,abah2012single,kosloff2014quantum,gelbwaser2015thermodynamics,uzdin2015equivalence,goold2016role,rossnagel2016single,vinjanampathy2016quantum,klaers2017squeezed,klatzow2019experimental,guff2018power} and heat diodes~\cite{ordonez2017quantum,kargi2019quantum,ronzani2018tunable} based on quantum systems require the ability to turn thermalization on and off as fast as possible, in order to maximize the heat flow (power) through the device.

\par

The basic questions in this context that may be partly elucidated by our study are: What are the criteria for the division of energy exchange between a quantum system and a bath into heat and work~\cite{niedenzu2016operation,niedenzu2018quantum}? What are the criteria for quantum thermalization enhancement? A growing trend is aimed at identifying quantum coherence/entanglement as a resource that can provide quantum advantages (supremacy) in a thermodynamic process~\cite{brandner2017universal}, so that the present study contributes to this trend. It extends previous works with featured baths consisting of entangled-qubit pairs~\cite{dag2019temperature} or triples~\cite{dag2016multiatom} as resources that can yield quantum advantage in either heat-transfer/thermalization or work.

\par

In Sec.~\ref{sec_master} we develop a master equation for repeated random interactions of the target qubit with multi-qubit clusters. Classification of multi-qubit persistent quantum coherences in the cluster according to their effect on the target qubit is presented in Sec.~\ref{sec_classification}. In Sec.~\ref{sec_qubit_temperature} we focus on the role of heat-exchange coherences (HEC) in the cluster and their preparation as regards the scaling of the target-qubit temperature and thermalization time. We discuss the results and their possible realizations in Sec.~\ref{sec_conclusions}.

\section{Master equation}\label{sec_master}

\par
\begin{figure}
  \centering
  \includegraphics[width=0.5\columnwidth]{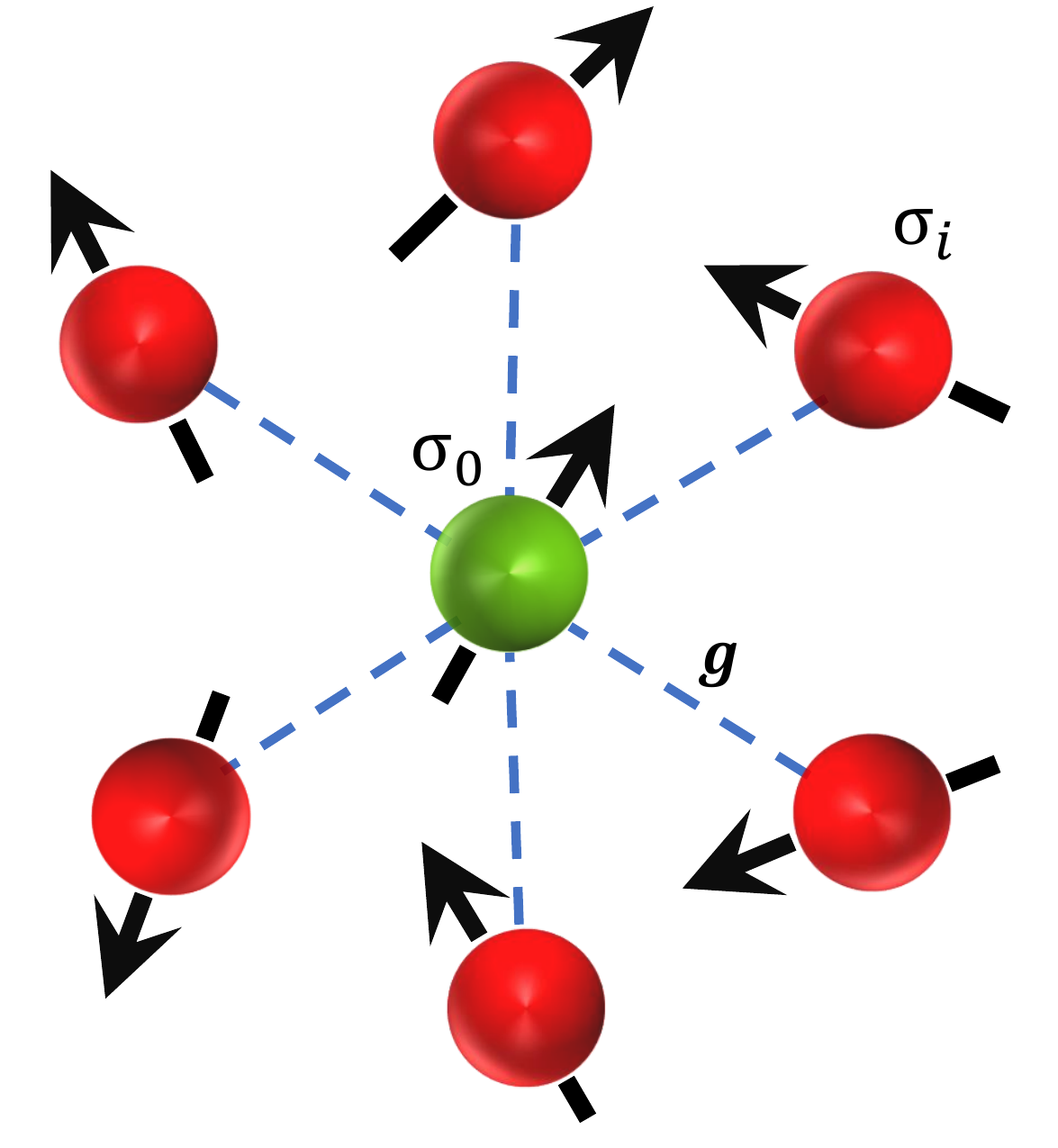}
  \caption{Interaction of a single (target) qubit with an atomic cluster (bath). The coupling $g$ of the target qubit to each bath qubit is supposed to be identical. After each interaction the state of the bath is reset (see text).}\label{fig_spins}
\end{figure}
\par

We adopt the commonly encountered central spin model~\cite{gaudin1976diagonalisation,prokofev2000theory,mukhopadhyay2017dynamics,guo2014dissipative} wherein a target qubit interacts at random times with a cluster of $N$ qubits, as shown in Fig.~\ref{fig_spins}. The total Hamiltonian reads $H=H_\mathrm{free}+H_\mathrm{int}$,
where
\begin{equation}\label{eq_H_free}
  H_\mathrm{free}=\frac{\hbar\omega_0}{2}\sigma _0^z+\frac{\hbar\omega_\mathrm{b}}{2}\sum_{i=1}^N\sigma _i^z
\end{equation}
is the free Hamiltonian of the target qubit and the $N$ bath qubits; $\sigma _i^z=|e_i\rangle\langle e_i|-|g_i\rangle\langle g_i|$ is the Pauli $z$ matrix for the $i$th qubit. For simplicity, we assume system-bath resonance, $\omega_0=\omega_\mathrm{b}$, and identical couplings such that
\begin{equation}\label{eq_H_int}
  H_{\mathrm{int}}=\hbar g \sum_{i=1}^N (\sigma _i^+ \sigma _0^- + \sigma _i^- \sigma _0^+)=\hbar g (J_+\sigma _0^-+J_-\sigma _0^+),
\end{equation}
where $\sigma_i^-=|g_i\rangle\langle e_i|$ and $\sigma_i^+=|e_i\rangle\langle g_i|$ are the individual lowering and raising operators and $J_{\pm}=\sum_{i=1}^{N}\sigma_i^{\pm}$ are the collective raising and lowering spin operators of the bath~\cite{breuerbook}, respectively.

\par

We consider the evolution of the system in the interaction picture with respect to the free Hamiltonian~\eqref{eq_H_free}. During the interaction time $\tau$, the system evolves unitarily according to the propagator $U(\tau)=\exp(-i H_{\mathrm{int}} \tau/\hbar)$ which, to second order in $g\tau$, reads
\begin{multline}\label{eq_U}
  U(\tau)\approx\mathds{1} -ig\tau(J_+\sigma _0^-+J_-\sigma _0^+) \\ 
  -\frac{(g\tau)^2}{2}(J_+J_-\sigma _0^-\sigma _0^++J_-J_+\sigma _0^+\sigma _0^-).
\end{multline}
Before each interaction, the system-bath density operator is supposed to be factorized, $\rho (t)=\rho _\mathrm{q}(t)\otimes \rho _\mathrm{b} $. This means that after each interaction the state of the bath is either reset or another replica of the bath is injected into the setup (see Sec.~\ref{sec_conclusions}). Since the target qubit interacts randomly with the bath qubits at a rate $p$, in a given time interval $\delta t$ the system either evolves according to Eq.~\eqref{eq_U} or remains invariant~\cite{dag2016multiatom,cakmak2017thermal,scullybook},
\begin{equation}
  \rho(t+\delta t)=p\delta tU(\tau)\rho(t)U^\dagger(\tau)+(1-p\delta t)\rho(t).
\end{equation}
In the limit $\delta t \rightarrow 0$, the master equation describing the dynamics of the target qubit is obtained by tracing out the degrees of freedom of the bath qubits, yielding~\cite{cakmak2017thermal}
\begin{align}\label{eq_master}
  \dot{\rho}_\mathrm{q}(t) & =\Tr_\mathrm{b}[p(U(\tau)\rho(t)U^\dagger(\tau)-\rho(t))] \notag\\
                          & =\frac{1}{i\hbar}[H_{\mathrm{eff}},\rho _\mathrm{q}]+\mathcal{L} _\mathrm{s} \rho_\mathrm{q} +\mathcal{L}_\mathrm{h} \rho_\mathrm{q},
\end{align}
where the effective Hamiltonian and the Lindblad operators are
\begin{subequations}\label{eq_H_L}
  \begin{align}
    H_{\mathrm{eff}}&=\hbar pg\tau (\lambda \sigma ^+ +\lambda ^* \sigma ^-) \label{eq_master_Heff}\\
    \mathcal{L} _\mathrm{s} \rho_\mathrm{q}&=\mu(\varepsilon  \sigma ^+ \rho _\mathrm{q} \sigma ^+ +\varepsilon ^* \sigma ^- \rho _\mathrm{q} \sigma ^-)\label{eq_master_Ls} \\
    \mathcal{L}_\mathrm{h} \rho_\mathrm{q}&=\frac{\mu r_\mathrm{d}}{2}\big(2\sigma ^- \rho _\mathrm{q} \sigma ^+-\sigma ^+ \sigma ^-  \rho _\mathrm{q}- \rho _\mathrm{q} \sigma ^+ \sigma ^-\big) \notag\\
                    &\quad+\frac{\mu r_\mathrm{e}}{2}\big(2\sigma ^+ \rho _\mathrm{q} \sigma ^--\sigma ^- \sigma ^+  \rho _\mathrm{q}- \rho _\mathrm{q} \sigma ^- \sigma ^+\big),\label{eq_master_Lh}
  \end{align}
\end{subequations}
where $\mu=p(g\tau)^2$. The other coefficients are defined in ~Table~\ref{table}.

\begin{table}
  \begin{tabular}{|P{1cm}|P{2.5cm}|P{3.5cm}|}
    \hline
    $\lambda$ & $\langle J_-\rangle$ & displaced bath \\ 
    $\varepsilon$ & $\langle J_-^2\rangle$ & squeezed bath \\
    $r _\mathrm{e}$ & $\langle J_+J_-\rangle$ & thermal bath (HEC)\\ 
    $r _\mathrm{d}$ & $\langle J_-J_+\rangle$ & thermal bath (HEC) \\
    \hline
  \end{tabular}
  \caption{The coefficients in the master equation~\eqref{eq_master} are related to expectation values of the collective bath spin. Their effect on the target qubit corresponds to an effective coherently-displaced, squeezed, or thermal bath.}\label{table}
\end{table}

The structure of the master equation~\eqref{eq_master} is in full analogy to the one obtained for the interaction of a single cavity field mode (harmonic oscillator) with a beam of $N=2$ or $N=3$ two-level atoms in a micromaser setup~\cite{dag2016multiatom,dag2019temperature}. 

\section{Effects of bath coherences on the target-qubit evolution}\label{sec_classification}

Table~\ref{table} plays an important role in the forthcoming discussions on the thermalization dynamics of the target qubit. It relates the coefficients of the master equation~\eqref{eq_master} to expectation values of the collective spin formed by the bath qubits. Each of these coefficients pertains to a different physical process:
\begin{itemize}
\item \emph{Displacement coherences:} These are coherences associated with the mean total polarization $\langle J_\pm \rangle$ of the collective bath spin that couples bath states differing by a single excitation. They occur in the effective Hamiltonian~\eqref{eq_master_Heff} that coherently drives the target-qubit transition with strength $|pg\tau\langle J_-\rangle|$ (analog to a laser drive~\cite{gardinerbook}). Hence, the bath clearly performs work on the system~\cite{niedenzu2016operation}.
\item \emph{Squeezing coherences:} These coherences are associated with the mean values of the two-excitation transitions in the bath, $\langle J_\pm^2\rangle$. They occur in the Liouvillian~\eqref{eq_master_Ls} which describes the interaction of the target qubit with an effective squeezed bath~\cite{breuerbook,cakmak2017thermal,dag2016multiatom}. 
\item \emph{Heat-exchange coherences (HEC):} These are coherences associated with $\langle J_\pm J_\mp\rangle$ that couple bath states with the same amount of excitations. They occur in the Liouvillian~\eqref{eq_master_Lh} which describes the de-excitation or the excitation of the target qubit with rates $\mu r_\mathrm{d}$ and $\mu r_\mathrm{e}$, respectively. This Liouvillian thus describes the interaction of the target qubit with an effective thermal bath.
\end{itemize}

\par

Thus, depending on the structure of the bath state, the target qubit may effectively be exposed to a coherently-displaced, squeezed or thermal environment (or any combination thereof). In contrast to earlier works~\cite{dag2016multiatom,dag2019temperature}, we here found the rapport between the structure and coefficients of the master equation~\eqref{eq_master} and many-body expectation values of the collective bath spin (Table~\ref{table}). The insights from Table~\ref{table} expose the intimate connection to superradiance, where the superradiance intensity of an ensemble of $N$ qubits is proportional to $\langle J_+J_-\rangle$~\cite{lambert2016superradiance}. Superradiance manifests itself by an $N$ times faster relaxation rate than a single qubit such that the intensity scales as $N^2$~\cite{breuerbook}. Analogous behavior will be shown here for collectively-enhanced thermalization.

\par


The insights from Table~\ref{table} allow us to answer for any $N$ (rather than only for $N\leq 3$ as in Ref.~\cite{dag2016multiatom}): How are the different coherences (displacement, squeezing and heat-exchange) distributed in the density matrix $\rho_\mathrm{b}$ of the bath?

\par

\begin{figure}
  \centering
  \includegraphics[width=0.95\columnwidth]{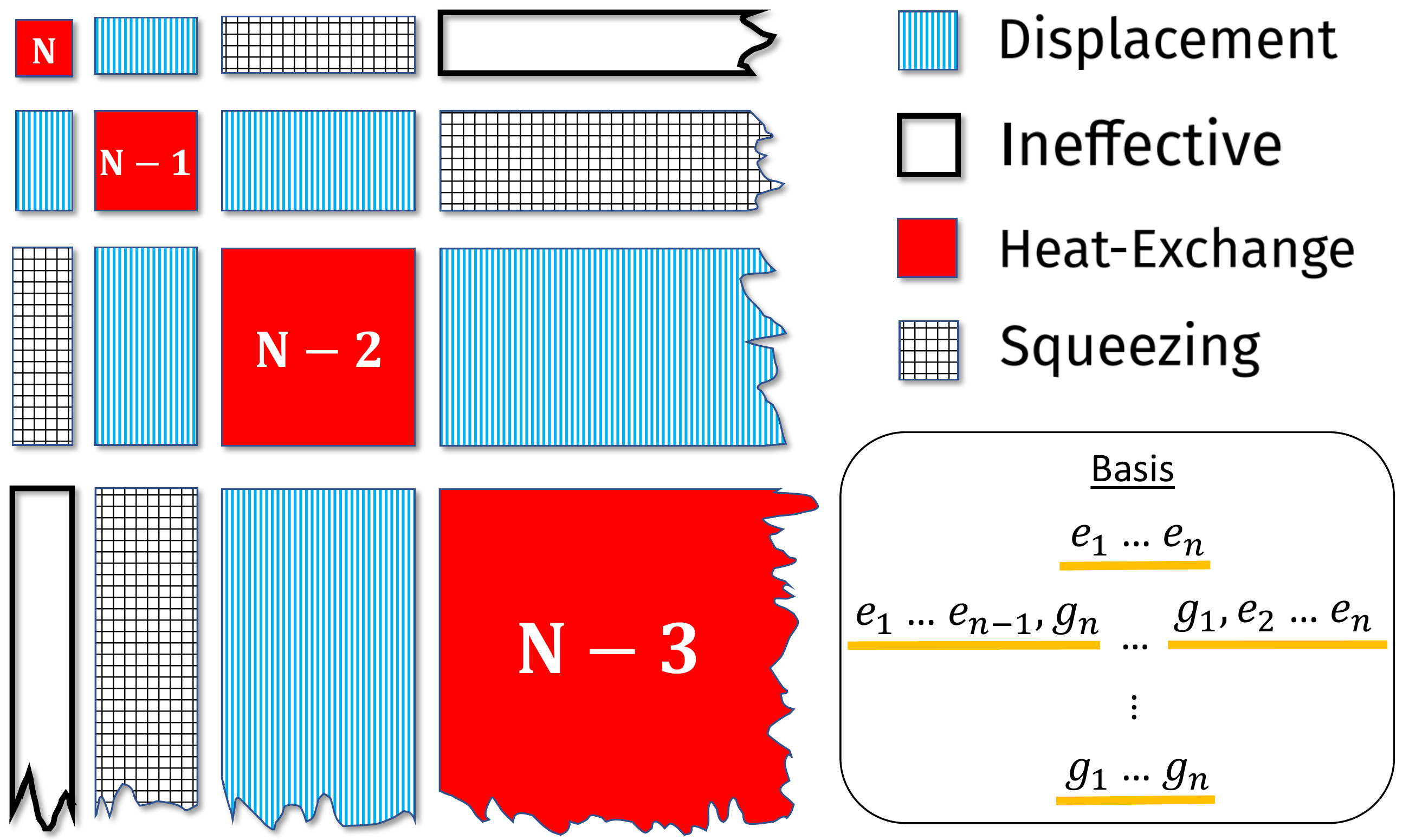}
  \caption{Density matrix of the bath qubits in the product-state basis. As described by the master equation~\eqref{eq_master}, different coherences in the bath have a specific role in the evolution of the target qubit. The main diagonal red squares are the states representing heat-exchange coherences (HEC) and populations that thermalize the target qubit by acting as an effective thermal bath. The numbers inside the red blocks indicate the number of excited qubits pertaining to the states in these blocks. Blue squares with vertical lines represent displacement coherences (corresponding to a coherently-displaced bath), and grey squares with a grid are the squeezing coherences (corresponding to an effective squeezed bath).}\label{fig_density}
\end{figure}

\begin{figure}
  \centering
  \includegraphics[width=0.6\columnwidth]{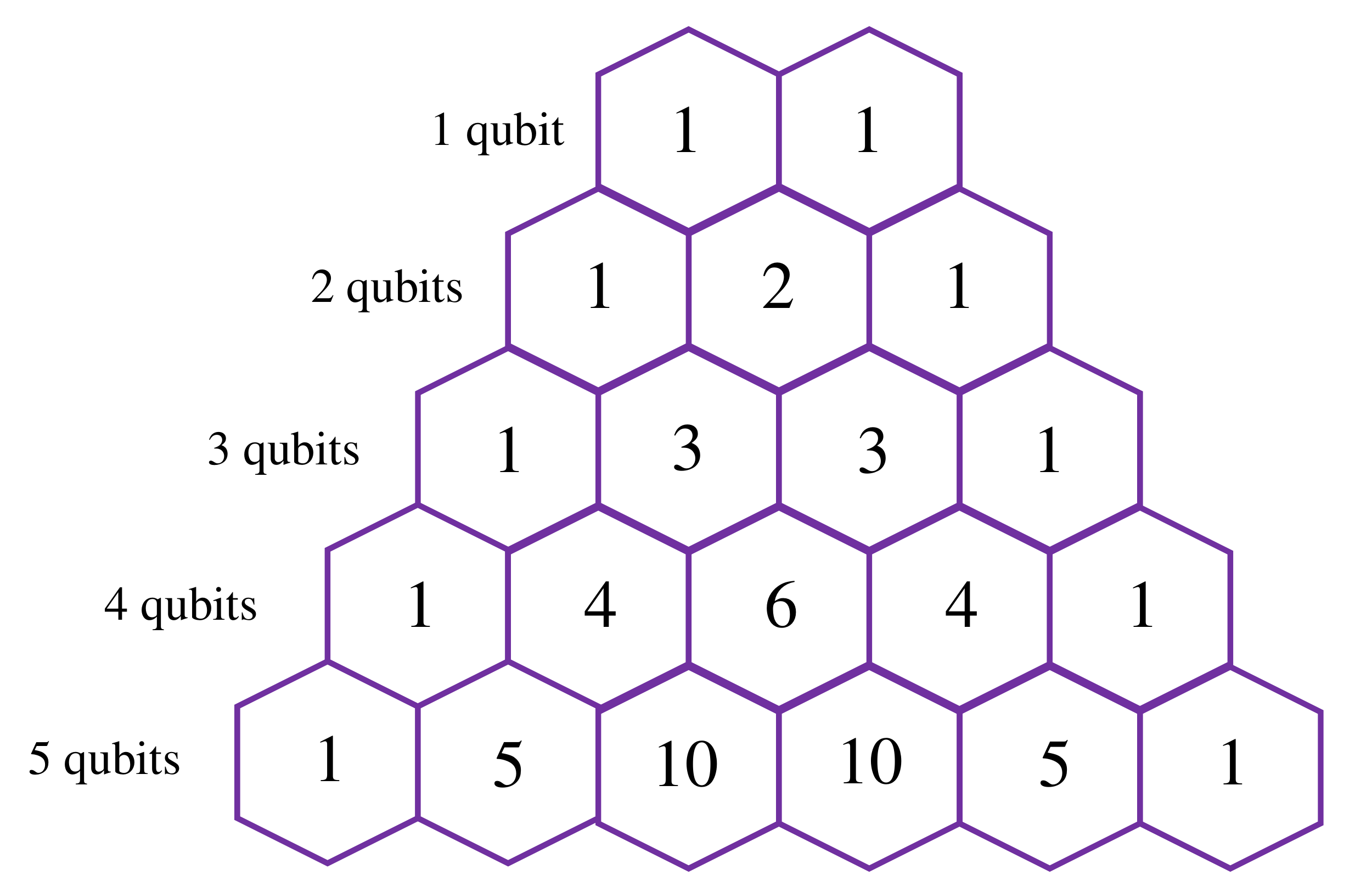}
  \caption{The sizes of the red HECs blocks in Fig.~\ref{fig_density} can be determined from Pascal's triangle; each cell in a line corresponds to $N,N-1,\dots,1,0$ excitations. The number in each cell (i.e., the size of the corresponding HEC block) is the sum of the two numbers in the two cells above it.}\label{fig_pascal}
\end{figure}

In the product-state basis the HECs are located in the off-diagonals of blocks of size
\begin{equation}\label{eq_p_k}
  p_k=\binom{N}{k}
\end{equation}
in the main diagonal of $\rho_\mathrm{b}$. Here $k=0, 1,\dots, N$ is the number of excitations of each state in block~$k$. These blocks are shown as solid red squares in Fig.~\ref{fig_density} and their size~\eqref{eq_p_k} can in a straightforward manner be determined from Pascal's triangle depicted in Fig.~\ref{fig_pascal}. The other coherences appearing in the master equation~\eqref{eq_master}, i.e., the displacement (squeezing) coherences, pertain to states differing by one (two) excitations and are shown in blue (grey) in Fig.~\ref{fig_density}.

\par

\begin{figure*}
  \centering
  \begin{center}
    \subfigure[2-qubit bath]{
      \label{fig:fig2a}
      \includegraphics[width=0.45\columnwidth]{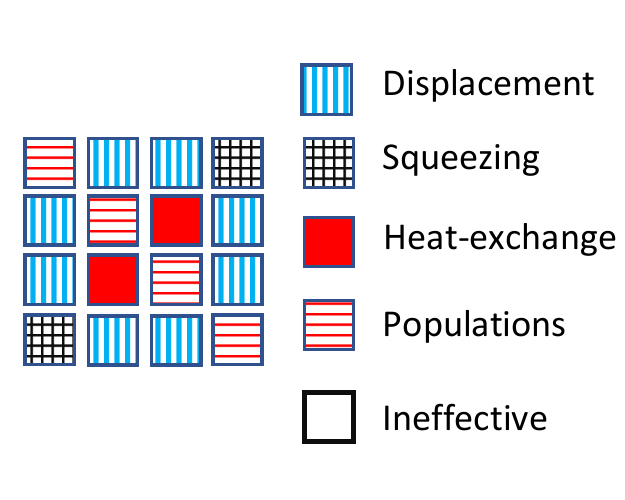}
    }
    \subfigure[3-qubit bath]{
      \label{fig:fig2b}
      \includegraphics[width=0.45\columnwidth]{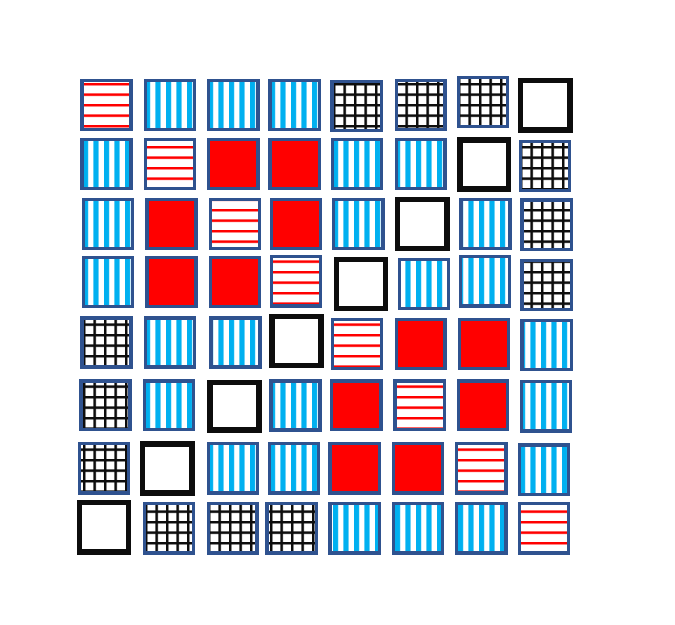}
    }
    \subfigure[4-qubit bath]{
      \label{fig:fig2c}
      \includegraphics[width=0.45\columnwidth]{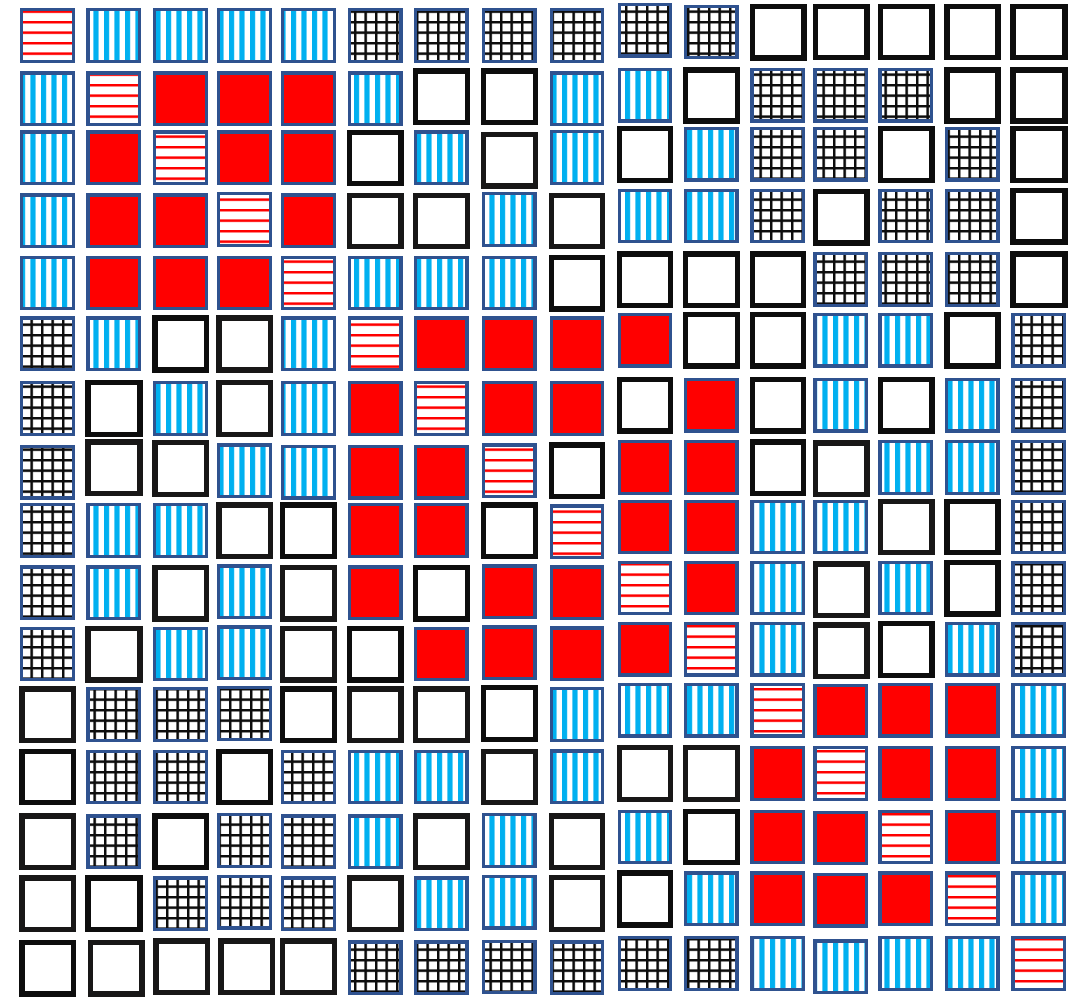}
    }
  \end{center}
  \caption{Density matrix of the bath (qubit clusters) from Fig.~\ref{fig_density} for different cluster sizes.}\label{fig_example}
\end{figure*}

The remaining coherences are ineffective as they do not contribute to the second-order Lindblad master equation~\eqref{eq_master}. Some of these ineffective coherences are scattered within the driving, squeezing and heat-exchange blocks (Fig.~\ref{fig_example}). The only ineffective coherences that appear in a HEC block correspond to the anti-diagonal of $\rho_\mathrm{b}$. These ineffective coherences also do not contribute to the generation of multi-partite correlations in a scenario where multi-qubit clusters repeatedly interact with independent qubits~\cite{daryanoosh2018quantum}.

\par
\par

\par
\par

\par
\par
      
\section{Qubit temperature and thermalization time}\label{sec_qubit_temperature}

In this work we wish to focus on the effects of HECs in the bath state. To this end we consider the scenario where only the thermal component~\eqref{eq_master_Lh} exists in the master equation~\eqref{eq_master} by setting $\lambda=\varepsilon=0$. We then ask: How do different bath states influence the coefficients $r_\mathrm{e}$ and $r_\mathrm{d}$ and hence the final temperature and the thermalization time of the target qubit?

\par

In this scenario, the elements of the target qubit density matrix evolve as~\cite{breuerbook}
\begin{subequations}\label{eq_rho_t}
  \begin{align}
    \rho_{eg}(t)&=\rho_{eg}(0)e^{-t/(2t_\mathrm{q})}\\
    \rho_{ge}(t)&=\rho_{eg}^*(t)\\
    \rho_{gg}(t)&=1-\rho_{ee}(t) \label{eq_rho_t_gg}\\
    \rho_{ee}(t)&=\frac{r_\mathrm{e}+c_0e^{-t/t_\mathrm{q}}}{r_\mathrm{e}+r_\mathrm{d}},\label{eq_rho_t_ee}
  \end{align}
\end{subequations}
where $c_0=r_\mathrm{d}\rho_{ee}(0)-r_\mathrm{e}\rho_{gg}(0)$. From Table~\ref{table}, the characteristic qubit thermalization time $t_\mathrm{q}^{-1}\coloneq\mu(r_\mathrm{e}+r_\mathrm{d})$ then evaluates to
\begin{equation}\label{eq_time}
  t_\mathrm{q}^{-1}=\mu\left(\langle J_+J_-\rangle+\langle J_-J_+\rangle\right).
\end{equation}
The steady-state of Eqs.~\eqref{eq_rho_t} is
\begin{equation}\label{eq_rho_ss}
  \rho_{\mathrm{ss}}=\frac{1}{r_\mathrm{e}+r_\mathrm{d}}
  \begin{pmatrix}
    r_\mathrm{e} & 0\\
    0 & r_\mathrm{d}
  \end{pmatrix},
\end{equation}
which for $r_\mathrm{d}>r_\mathrm{e}$ corresponds to a thermal state with the well-defined temperature
\begin{equation}\label{eq_temp}
 T_\mathrm{q}=-\frac{\hbar \omega_0}{\kB \ln \left(\frac{r_\mathrm{e}}{r_\mathrm{d}}\right)}=-\frac{\hbar \omega_0}{\kB \ln \left(\frac{\langle J_+J_-\rangle}{\langle J_-J_+\rangle}\right)}.
\end{equation}
It can be seen that both the final temperature~\eqref{eq_temp} of the qubit and its thermalization time~\eqref{eq_time} explicitly depend on the many-body quantum state of the bath, as will be illustrated in several generic cases. We note that the diagonal character of the state~\eqref{eq_rho_ss} straight-forwardly allows the attribution of the temperature~\eqref{eq_temp} to this state. While one can always attribute a temperature to a diagonal two-level system, we are confident that this is also possible for an extension to higher-dimensional target systems, in analogy to our previous studies~\cite{dag2016multiatom,dag2019temperature} of a resonator mode (harmonic oscillator) that is driven by clusters of correlated or entangled atoms in a micromaser setup.

\subsection{Incoherent bath qubits}

First, we consider the simplest case in which every bath qubit is prepared in an incoherent mixture
\begin{equation}\label{eq_rho_i_mix}
  \rho_i^\mathrm{mix}=p_g\proj{g_i}+p_e\proj{e_i}
\end{equation}
of its excited and ground state with probabilities $p_{e}$ and $p_{g}>p_e$, respectively, such that
\begin{equation}\label{eq_rhob_mix}
  \rho_\mathrm{b}=\bigotimes_{i=1}^N \rho_i^\mathrm{mix}.
\end{equation}
The states~\eqref{eq_rho_i_mix} may have been individually prepared by a thermal environment at temperature $T$, in this case $p_e=p_g \exp[-\hbar\omega_0/(\kB T)]$.

\par

For the state~\eqref{eq_rhob_mix} we find $r_\mathrm{e}=Np_e$ and $r_\mathrm{d}=Np_g$ such the target qubit temperature~\eqref{eq_temp} evaluates to
\begin{equation}
  T_\mathrm{q}^\mathrm{mix}=-\frac{\hbar\omega_0}{\kB\ln\left(\frac{p_e}{p_g}\right)}=T,
\end{equation}
which is the temperature of an individual bath qubit. The corresponding thermalization time~\eqref{eq_time} of the target qubit is
\begin{equation}\label{eq_time_mix}
  t_\mathrm{q}^\mathrm{mix}=\frac{1}{\mu N}.
\end{equation}
Hence, the target qubit thermalizes to the temperature of each bath qubit, but $N$-times faster.
The decrease of the thermalization time with $N^{-1}$ shows that no quantum advantage applies here.

\par

A similar result was obtained in~\cite{turkpence2019tailoring} for a bosonic single-mode field (oscillator) instead of the target qubit.

\subsection{Thermally-prepared HECs}

Next we consider a correlated state of the bath qubits with non-vanishing HECs. Such a state may either be prepared by \emph{thermal} or \emph{non-thermal} means. We first dwell on the former case which may be realized by collectively (rather than individually as above) coupling the $N$ bath qubits to a thermal photon environment, thus giving rise to the master equation~\cite{breuerbook,cakmak2017thermal,shammah2018open}
\begin{multline}\label{eq_master_bath}
  \dot{\rho}_\mathrm{b}=\frac{\gamma_0}{2}(\bar{n}+1)\left(2J_-\rho_\mathrm{b} J_+-J_+J_-\rho_\mathrm{b}-\rho_\mathrm{b} J_+J_-\right)\\
  +\frac{\gamma_0}{2}\bar{n}\left(2J_+\rho_\mathrm{b}J_--J_-J_+\rho_\mathrm{b}-\rho_\mathrm{b}J_-J_+\right),
\end{multline}
where $\gamma_0$ is the single-atom spontaneous emission rate and $\bar{n}$ the mean number of photons in mode $\omega_\mathrm{b}$ of the environment at temperature $T$. Assuming that all the bath qubits were initially prepared in their ground state, the dynamics can then only populate the fully-symmetric Dicke states~\cite{breuerbook}, which correspond to the HECs depicted in Fig.~\ref{fig_density} (see Appendix~\ref{app_dicke}). The steady state of the bath qubits then reads 
\begin{equation}\label{eq_rhob_hec_thermal}
  \rho_\mathrm{b}=
  \begin{pmatrix}
    D_{N} & 0 &\dots & 0 &0\\
    0 & D_{N-1} & \dots & 0 & 0\\
    \vdots & \vdots & \ddots & \vdots & \vdots \\
    0 & 0 & \dots &  D_1 & 0\\
    0 & 0 & \dots & 0 & D_0
  \end{pmatrix},
\end{equation}
with the $(p_k\times p_k)$-dimensional matrices $D_k=d_k U_k$ [cf.\ Eq.~\eqref{eq_p_k}]. Here $U_k$ denotes the matrix whose elements are all $1$ (which means that only the fully-symmetric Dicke states are populated) and the coefficients $d_k$ evaluate to (see Appendix~\ref{app_dicke})
\begin{equation}
  d_k=\frac{(1-r)r^k}{(1-r^{N+1})p_{k}},
\end{equation}
where $r\coloneq\bar{n}/(\bar{n}+1)$.

\par

We now consider the thermally-prepared block-diagonal state~\eqref{eq_rhob_hec_thermal} as the bath for the target qubit. We then find $r_\mathrm{e}/r_\mathrm{d}=r$, which only depends on $\bar{n}$ (the explicit expressions are given in Appendix~\ref{app_coefficients}). Consequently, the target qubit attains the same temperature $T_\mathrm{q}^\mathrm{HEC}=T$ [Eq.~\eqref{eq_temp}] as the environment that was used to prepare the coherent bath qubits in the state~\eqref{eq_rhob_hec_thermal}. Hence, such thermally-prepared bath clusters do not show any quantum superiority in the final temperature of the target qubit.

\par

The thermalization time~\eqref{eq_time} of the target qubit evaluates to
\begin{equation}\label{eq_time_hec}
  t_\mathrm{q}^\mathrm{HEC}=\frac{1}{\gamma_\mathrm{eff}(2\bar{n}+1)},
\end{equation}
where $\gamma_{\rm eff}\coloneq\mu r_\mathrm{e}/\bar{n}=\mu r_\mathrm{d}/(\bar{n}+1)$ is the effective spontaneous emission rate that depends on the number of qubits (see Appendix~\ref{app_coefficients}) and, basically, increases linearly with $N$.

\par

Hence, regardless of whether the bath consists of $N$ independent thermal spins [Eq.~\eqref{eq_time_mix}] or a thermal collective spin [Eq.~\eqref{eq_time_hec}], the thermalization rate is sped up linearly in $N$ in either case, which is also expected in a classical setting. Furthermore, the temperature attained by the target qubit always coincides with the environment temperature used to prepare the bath. This leads to the conclusion that a \emph{non-thermal} generation of the HECs is required in order to exhibit possible quantum advantages of the bath.

\subsection{Non-thermally prepared HECs: Dicke bath}

We now consider a different type of $N$-qubit bath, namely, a fully-symmetric Dicke state with $k$ excitations, such that only one of the blocks in the state~\eqref{eq_rhob_hec_thermal} is populated,
\begin{equation}\label{dmcenter}
\rho_\mathrm{b}=
\begin{pmatrix}
0 & \dots &\dots & 0 \\
\vdots & \ddots & \dots & 0 \\
\vdots & \vdots & D_k & \vdots  \\
0 & 0 & \dots &  0 \\
\end{pmatrix},
\end{equation}
where normalization now implies $D_k=U_k/p_k$. A possible preparation method of such states has been recently proposed in Ref.~\cite{cakmak2019robust}. For the state~\eqref{dmcenter} we find
\begin{subequations}
  \begin{align}
    r_\mathrm{e}&=\langle J_+J_-\rangle=k(N-k+1)\\
    r_\mathrm{d}&=\langle J_-J_+\rangle=(k+1)(N-k).
  \end{align}
\end{subequations}
Thus, the steady-state temperature of the target qubit is
\begin{equation}\label{eq_Tq_k}
  T_\mathrm{q}^\mathrm{D}=-\frac{\hbar\omega_0}{\kB \ln\left[\frac{k(N-k+1)}{(k+1)(N-k)}\right]},
\end{equation}
which is positive as long as $r_\mathrm{d}>r_\mathrm{e}$, yielding the non-inversion condition
\begin{equation}
  k\leq \lceil N/2\rceil-1.
\end{equation}
Namely, there must always be more bath qubits in the ground state than in the excited state.

\par

The temperature~\eqref{eq_Tq_k} depends on the number of excitations $k$ and the number of bath qubits $N$. For large $N$ and large $k$ (such that $r_\mathrm{d}/r_\mathrm{e}\gtrsim 1$) the Laurent expansion yields
\begin{equation}\label{eq_T_dicke}
  \frac{\kB T_\mathrm{q}^\mathrm{D}}{\hbar\omega_0}\approx\frac{(k+1)(N-k)}{(k+1)(N-k)-k(N-k+1)}-\frac{1}{2}.
\end{equation}
Since the temperature~\eqref{eq_Tq_k} depends on $N$, also the target qubit's entropy $S=-\kB\Tr(\rho_\mathrm{ss}\ln\rho_\mathrm{ss})$ explicitly depends on the amount of bath qubits. In the high-temperature limit of Eq.~\eqref{eq_T_dicke} this dependence, however, becomes marginal as $S=k_\mathrm{B}\ln 2-\mathcal{O}\left(\middle[1-r_\mathrm{e}/r_\mathrm{d}]^2\right)$, consistent with the fact that the qubit levels are almost equally populated.

\par

If we consider a block containing a relatively small number of excitations, i.e., away from the central block, for example $k=N/4$ (assuming $N$ is an integer multiple of $4$), we find
\begin{equation}\label{eq_T_dicke_linear}
  \frac{\kB T_\mathrm{q}^\mathrm{D}}{\hbar\omega_0}\approx 1+\frac{3N}{8}.
\end{equation}
Hence, in this case, the scaling of the temperature is linear with the number of qubits in the bath cluster. On the other hand, increasing the number of excitations to $k=N/2-1$, i.e., choosing the closest allowed block to the central one, yields
\begin{equation}\label{eq_T_dicke_quadratic}
  \frac{\kB T_\mathrm{q}^\mathrm{D}}{\hbar\omega_0}\approx \frac{N^2+2N}{8}-\frac{1}{2}.
\end{equation}
Here we observe a clear \emph{quadratic scaling} with $N$. Hence, by varying the number of excitation in the bath, the temperature scaling may exhibit a strong \emph{collective quantum advantage} by growing quadratically with the bath-ensemble size $N$. This constitutes the second main result of this work.

\par

\begin{figure}
  \centering
  \includegraphics[width=.9\columnwidth]{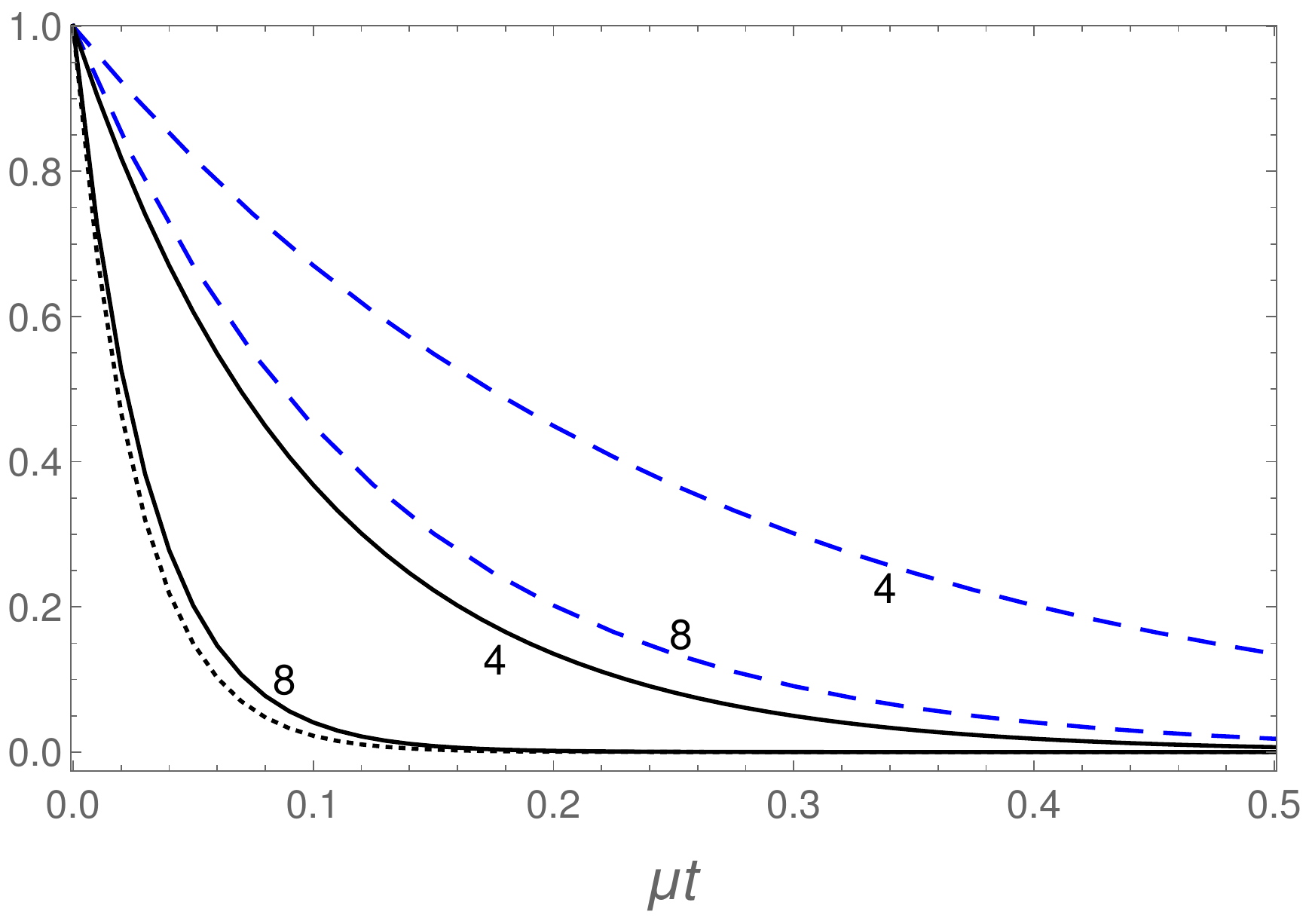} 
  \caption{Decay term $\exp(-t/t_\mathrm{q})$ [Eq.~\eqref{eq_rho_t_ee}] as a function of the scaled time $\mu t$ when the $N$-qubit bath is in a fully-symmetric Dicke state with $k=N/4$ excitations (black-solid) or in a mixed state (blue-dashed) for $N=4$ and $N=8$, respectively. The respective thermalization times are $t_\mathrm{q}^\mathrm{mix}$ [Eq.~\eqref{eq_time_mix}] and $t_\mathrm{q}^\mathrm{D}(k)$ [Eq.~\eqref{eq_time_dicke}]. The black-dotted line is for a fully-symmetric Dicke state of $N=8$ spins with $k=N/2-1=3$ excitations.}\label{fig_thermalization}
\end{figure}

This favorable scaling is also manifest in the corresponding thermalization time~\eqref{eq_time}, which for the Dicke state~\eqref{dmcenter} is found to be
\begin{equation}\label{eq_time_dicke}
  t_\mathrm{q}^\mathrm{D}(k)=\frac{1}{\mu(N+2kN-2k^2)}.
\end{equation}
Hence, contrary to the previous cases of thermally-prepared bath states, the thermalization time may be sped-up quadratically in $N$. For the examples above, $k=N/4$ and $k=N/2-1$, we find
\begin{subequations}
  \begin{equation}
    t_\mathrm{q}^\mathrm{D}(N/4)=[\mu(N+3N^2/8)]^{-1}
  \end{equation}
  and
  \begin{equation}
    t_\mathrm{q}^\mathrm{D}(N/2-1)=[\mu(N+N^2/2-2)]^{-1},
  \end{equation}
\end{subequations}
respectively. Therefore, coherences in the $N$-qubit bath prepared by \emph{non-thermal} means allow us to thermalize the the target qubit faster (optimally with $N^{-2}$ dependence) than their thermal (individual or collective) counterparts (optimally with $N^{-1}$ dependence), as shown in Fig.~\ref{fig_thermalization}.

\par
\par

\par

In principle, the bath state~\eqref{dmcenter} could also be used to cool the target qubit below its initial temperature by adjusting $N$ and $k$. The $N^2$ speedup, however, would only occur if $k\sim N$, which also implies larger temperatures. An option to optimize cooling in terms of final temperature and cooling time would be to investigate more general bath states, similar to Ref.~\cite{dag2019temperature}.

\par

Finally, we now investigate whether the observed quantum advantage in the target qubit temperature is present throughout the entire dynamics or only at steady state when interacting with the Dicke bath~\eqref{dmcenter}. If the target qubit is prepared in a state devoid of coherences, we can define its time-dependent temperature as $T_q(t):=(\hbar\omega_0/k_B)\ln[\rho_{gg}(t)/\rho_{gg}(t)]^{-1}$. For the initial state being the ground state ($\rho_{eg}(0)=\rho_{ee}(0)=0$ and $\rho_{gg}(0)=1$), Eqs.~\eqref{eq_rho_t_gg} and~\eqref{eq_rho_t_ee} yield
\begin{equation}\label{temperature_time}
  T_q(t)=\frac{\hbar\omega_0}{k_B}\left[\ln\left(\frac{1+r_d/r_e}{1-e^{-t/t_q}}-1 \right) \right]^{-1}.
\end{equation}

\par

\begin{figure}
  \centering
  \includegraphics[width=0.98\columnwidth]{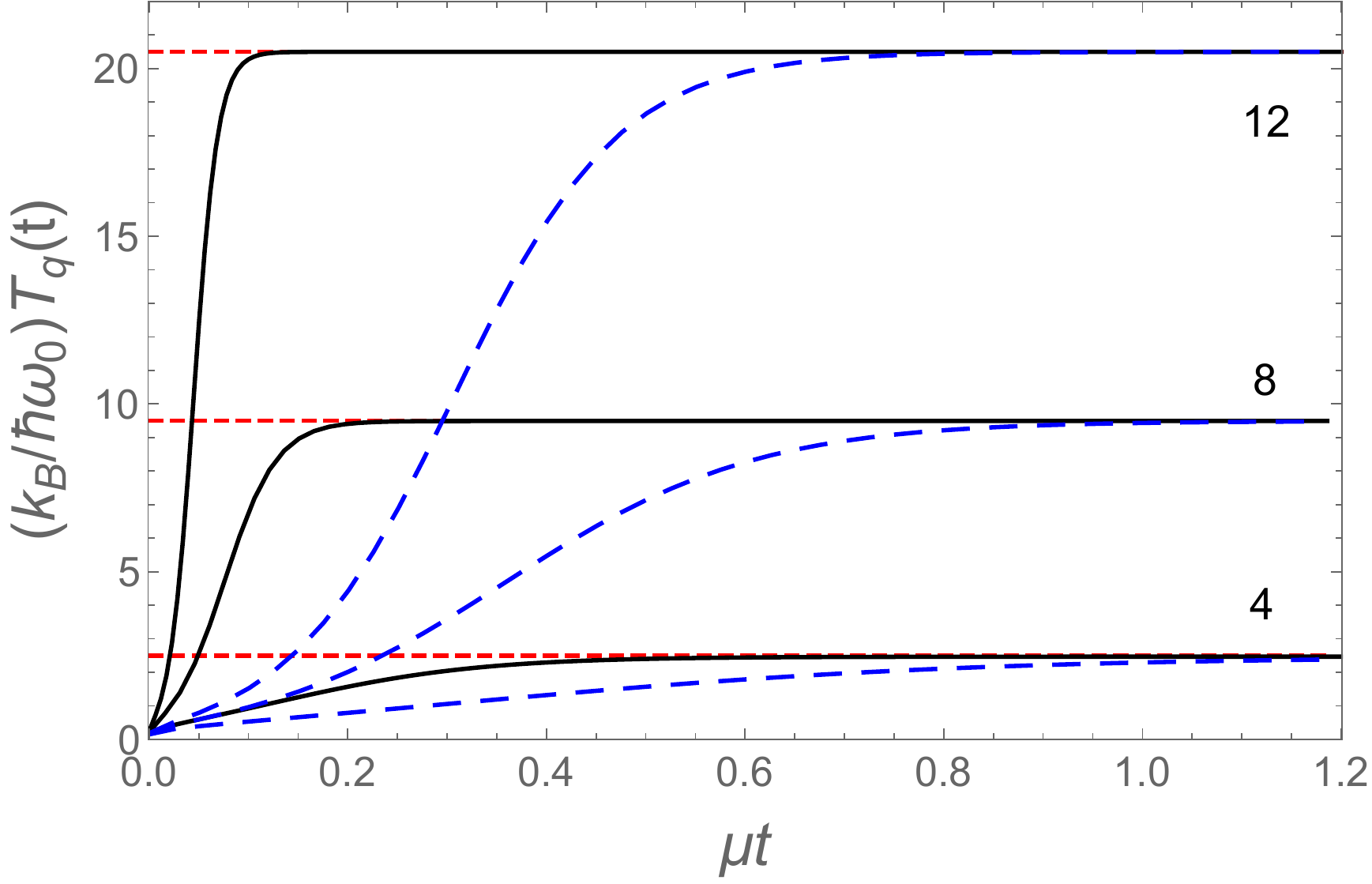} 
  \caption{Temperature~\eqref{temperature_time} of the target qubit as a function of the scaled time $\mu t$. Initially in its ground state (zero temperature), the qubit evolves to a thermal state defined by the $N$-qubit bath. Black solid lines: Dicke bath [Eq.~\eqref{dmcenter}] with $k=N/2-1$ excitations; blue dashed lines: Incoherent bath state~\eqref{eq_rhob_mix} with $T_\mathrm{q}^\mathrm{mix}=T_\mathrm{q}^\mathrm{D}$. Horizontal red dashed lines: Temperature estimate~\eqref{eq_T_dicke_quadratic} for $N=\{4,8,12\}$, depicting the quadratic dependence on $N$.}\label{fig_thermalization_temperature}
\end{figure}

The time-dependent temperature~\eqref{temperature_time} is shown in Fig.~\ref{fig_thermalization_temperature} for different sizes of incoherent and Dicke baths, respectively. The quantum advantage of the Dicke bath is present at all times, meaning that the target qubit temperature always surpasses the one generated by an incoherent bath.

\par
\par

\section{Conclusions}\label{sec_conclusions}

We have investigated a collisional model wherein a single target qubit repeatedly interacts with clusters of $N$ bath qubits. We have observed that the evolution of the target qubit strongly depends on the quantum state of the bath. We have established the dependence of the qubit evolution on the first and second moments (linear and quadratic means) of the collective spin of the bath (Table~\ref{table}), and showed that according to the class of multi-qubit coherences in the bath clusters, the bath can thermalize, cause its coherent drive or simulate squeezed-bath effects on the target qubit.

\par

Focusing on the case of thermalization dynamics of the target qubit, we have considered different types of coherent and incoherent bath clusters; either thermally or non-thermally prepared. We have shown that thermally-prepared bath states cause thermalization of the target qubit to the temperature of the environment that was used to create them. Whether the bath consists of $N$ independent thermal qubits [Eq.~\eqref{eq_rhob_mix}] or thermally prepared coherent qubits [Eq.~\eqref{eq_rhob_hec_thermal}], both cases show a decrease in the target qubit thermalization time proportional to $N^{-1}$.

\par

On the other hand, non-thermally-prepared bath states with only one non-zero Dicke block yield either linear [Eq.~\eqref{eq_T_dicke_linear}] or quadratic [Eq.~\eqref{eq_T_dicke_quadratic}] scaling of the target qubit temperature~\eqref{eq_T_dicke} with the number of qubits in the bath ensemble, depending on the number of excitations in the chosen block. The largest quantum advantage corresponds to quadratic scaling $T\propto N^2$. Likewise, the relaxation time may decrease quadratically, $t_\mathrm{q}\propto N^{-2}$. These results are the main contribution of the present work since they explicitly show a broad control in the thermalization temperature of a target system and its time, using a structured quantum bath, which otherwise would not be possible by utilizing classical resources. This $N^2$ quantum advantage obtained here is reminiscent of superradiance. Yet, whereas superradiance occurs for the collective interaction of an ensemble of qubits with a common bath~\cite{breuerbook}, here it is the bath that exhibits collective behavior such that the constituents of the bath interact with the single target qubit at the same time. It is interesting to note that the bath states that depict a quantum advantage, i.e., the Dicke bath state~\eqref{dmcenter}, are multipartite entangled $W$-type states whereas the incoherent and thermally prepared HEC states without quantum advantage are separable. It would be interesting to further investigate the role of multipartite entanglement on thermalization via multi-qubit collistions in the spirit of Refs.~\cite{dag2019temperature,daryanoosh2018quantum}.

\par

The Hamiltonian~\eqref{eq_H_int} effectively describes various systems, e.g., nuclear spin baths in quantum dots~\cite{chekhovich2013nuclear,jing2018decoherence}, nitrogen-vacancy (NV) centers~\cite{reinhard2012tuning}, nuclear magnetic resonance systems~\cite{pal2018temporal}, microcavities hosting multi-atom systems~\cite{hartmann2007effective,wan2009tunable} or multiple superconducting qubits~\cite{niskanen2007quantum,cakmak2019robust}, and molecular nanomagnets~\cite{ardavan2007will,chiesa2014molecular}.

\par

However, it is not straightforward in the above-mentioned scenarios to mimic the effect of repeated impulsive interactions of a target qubit and non-thermally prepared $N$-qubit bath states. The required protocol would then consist of multiple steps, each step involving: controlled preparation of the $N$-qubit (entangled) bath state, abrupt on- and off-switching of its interaction with the target qubit followed by resetting of the $N$-qubit bath to its initial state.

\par

There is, however, a simpler and more natural scenario where the described process can be implemented: It concerns a cold-atom cloud with, upon absorbing a few quanta, attains a superradient (symmetric) Dicke state~\cite{mazets2007multiatom,araujo2016superradiance,guerin2016subradiance}. Following such state preparation, an impurity atom passing near or within this cloud will, to a good approximation, realize the described model.

\par

A future extension to non-Markovian baths~\cite{mccloskey2014nonmarkovianity,cakmak2017nonmarkovianity} may be interesting.

\section*{Acknowledgements}

\"{O}.\,E.\,M.\ acknowledges support by TUBITAK (Grant No.\ 116F303) and by the EU-COST Action (CA15220). W.\,N.\ acknowledges support from an ESQ fellowship of the Austrian Academy of Sciences (\"OAW). G.\,K. acknowledges support from ISF, SAERI and DFG. \"{O}.\,E.\,M.\ would like to thank C.\,B.\,Da\u{g} and F.\,Ozaydin for initial discussions on the problem.

\appendix

\section{Thermal preparation of HECs}\label{app_dicke}

\begin{figure}
  \centering
  \includegraphics[width=.9\columnwidth]{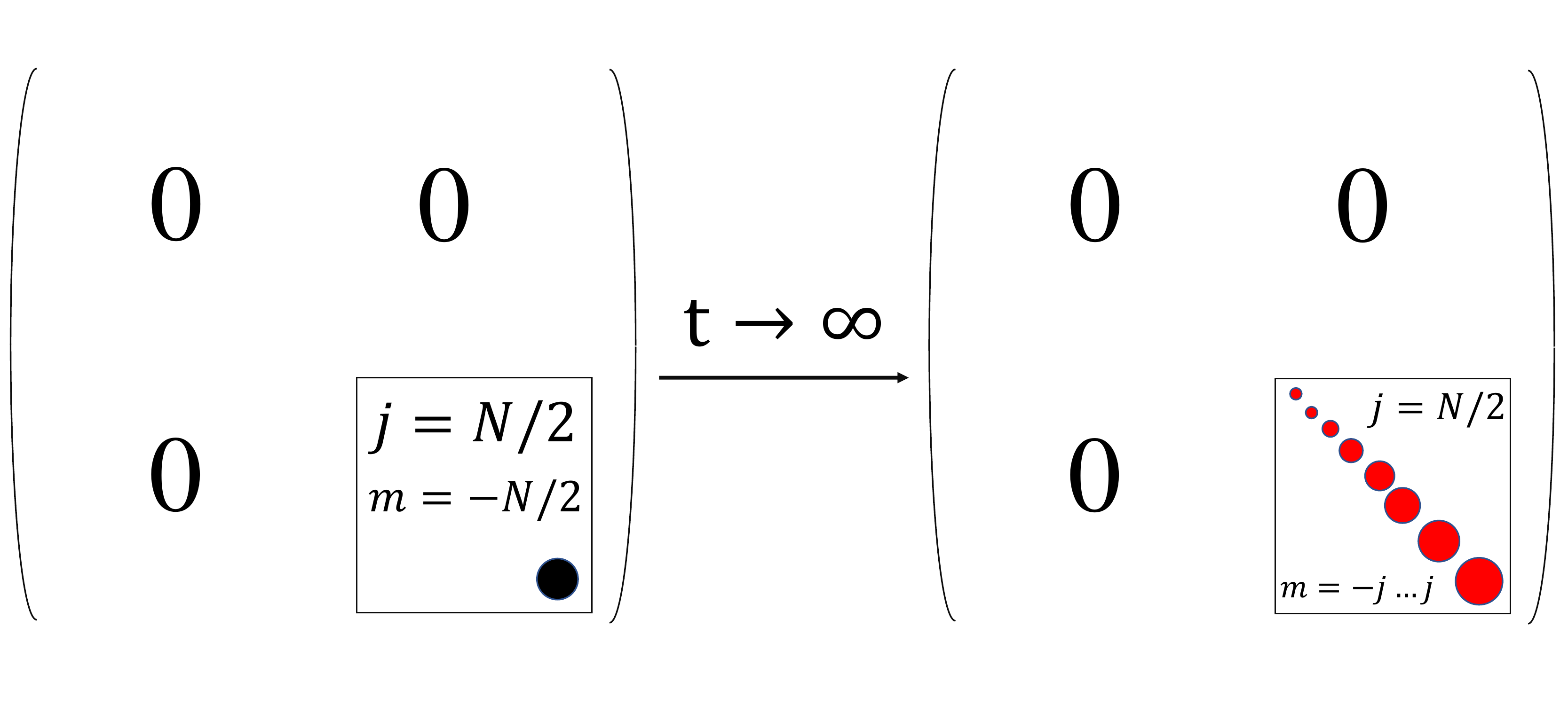}
  \caption{Time evolution in the Dicke basis.}\label{lindblad}
\end{figure}

We may express the density matrix $\rho_\mathrm{b}$ of the bath spins in either the product-state basis or the total-spin (Dicke) basis $|j,m\rangle$ where $j=0,\dots,N/2$ ($j=1/2,\dots,N/2$) for even (odd) $N$ and $m=-j,\dots,j$~\cite{breuerbook}. Each $m$ corresponds to a distinct HEC block in the product-state basis. If, as in the main text, the atoms are initially prepared in their ground state $|g_1,g_2,\dots,g_N\rangle\equiv |j=N/2,m=-N/2\rangle$, the dynamics governed by the master equation~\eqref{eq_master_bath} do not mix the different $j$, i.e.,
\begin{multline}
  \rho_\mathrm{b}(0)=\proj{\frac{N}{2},-\frac{N}{2}} \longrightarrow \\\rho_\mathrm{b}(t)=\sum\limits_{m=-N/2}^{N/2}\rho_{m}(t)\proj{\frac{N}{2},m}.
\end{multline}
The time evolution according to Eq.~\eqref{eq_master_bath} then thermally populates every $\ket{N/2,m}$ such that in steady state (see Fig.~\ref{lindblad})
\begin{equation}
  \rho_m=\frac{1}{Z}\exp\left(-\frac{m\hbar\omega_0}{\kB T}\right)
\end{equation}
with
\begin{equation}
  Z=\sum_{m=-N/2}^{N/2}\exp\left(-\frac{m\hbar\omega_0}{\kB T}\right).
\end{equation}

\par
\par

\par
\begin{figure}
  \centering
  \includegraphics[width=.9\columnwidth]{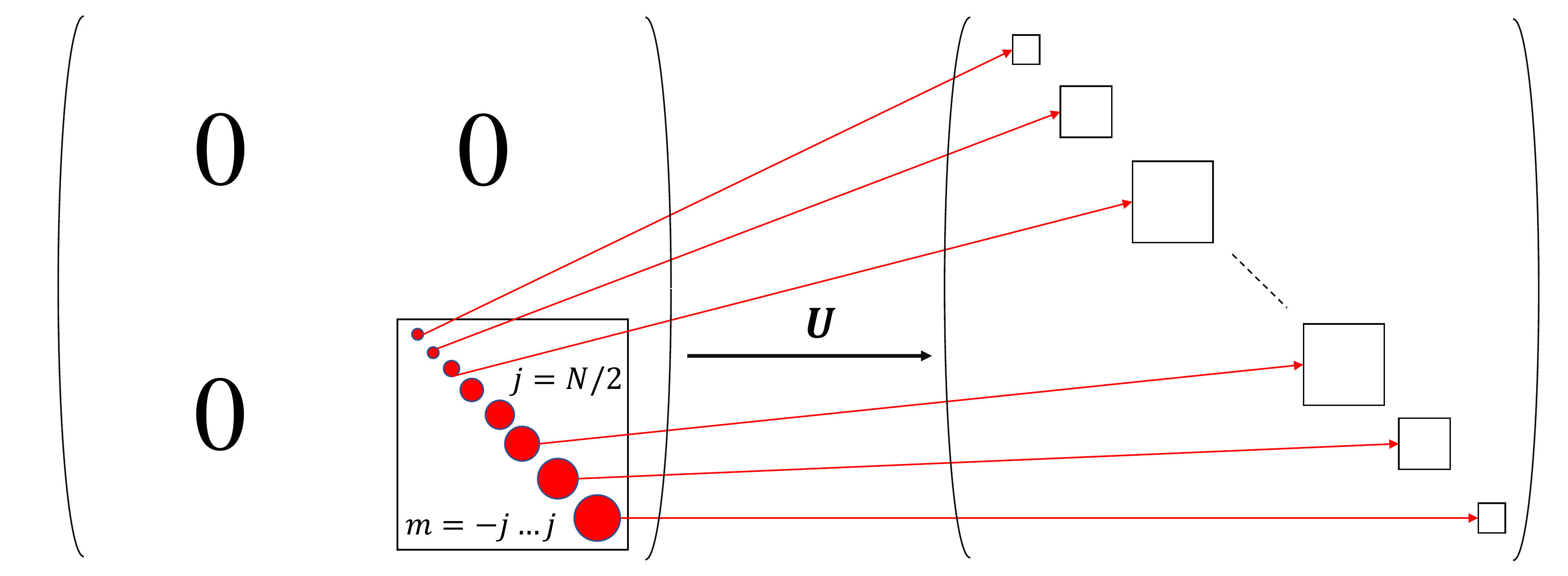}
  \caption{Transformation from the Dicke basis to the product-state basis.}
  \label{fig_transform}
\end{figure}
\par

Upon transforming to the product-state basis by a unitary transformation $U$, the population $\rho_m$ of every fully-symmetric state $|N/2, m\rangle$ is transferred to the corresponding HEC block (Fig.~\ref{fig_transform}). Each such block has dimensionality $(p_k\times p_k)$ [Eq.~\eqref{eq_p_k}] and may be written $D_k=d_k U_k$, where $U_k$ denotes the matrix whose elements are all $1$. Hence, the population ratio $\rho_{m+1}/\rho_m$ is directly reflected in the ratio of the traces of two consecutive blocks,
\begin{equation}
\frac{ \Tr[D_{m+1}]}{ \Tr[D_{m}]}=\frac{\rho_{m+1}}{\rho_m}=\exp\left(-\frac{\hbar\omega_0}{\kB T}\right)=\frac{\bar{n}}{\bar{n}+1}\eqcolon r,
\end{equation}
From the normalization condition
\begin{equation}\label{trac}
  1=\Tr(\rho_\mathrm{b})=\sum_{k=0}^{N} \Tr[D_k]=\sum_{k=0}^{N} r^k\Tr[D_0]
\end{equation}
follows
\begin{equation}
  \Tr[D_0]=\frac{1-r}{1-r^{N+1}}.
\end{equation}
Since $D_k=d_kU_k$, we have $d_k=\Tr[D_k]/\Tr[U_k]$. Using $\Tr[D_k]=r^k\Tr[D_0]$ and $\Tr[U_k]=p_k$ then yields
\begin{equation}
  d_k=\frac{(1-r)r^k}{(1-r^{N+1})p_{k}}.
\end{equation}

\section{Explicit form of the rates}\label{app_coefficients}

For the state~\eqref{eq_rhob_hec_thermal} the coefficients in Table~\ref{table} evaluate to~\cite{cakmak2017thermal} 
\begin{align}\label{re_hec}
  r_\mathrm{e} &= \Tr[J_-\rho_\mathrm{b}J_+]=\sum_{k=0}^{N}d_k\Tr[J_-U_k J_+] \notag\\
               &= \sum_{k=1}^{N}d_k (N-k+1)^2 p_{k-1} \notag\\
               &= \sum_{k=1}^{N}\frac{(1-r) r^{k}k(N-k+1)}{1-r^{N+1}}
\end{align}
and
\begin{align}\label{rd_hec}
  r_\mathrm{d} &= \Tr[J_+\rho_\mathrm{b}J_-]=\sum_{k=0}^{N}d_k\Tr[J_+U_k J_-] \notag\\
              &= \sum_{k=0}^{N-1}d_k(k+1)^2 p_{k+1} \notag\\
              &= \sum_{k=1}^{N}\frac{(1-r)r^{k-1}k(N-k+1)}{1-r^{N+1}}.
\end{align}

\end{document}